\documentclass[shortnote%
 	,twocolumn
 	]{jpsj3}
\usepackage{txfonts}
\title{Adiabatic Theorem for Discrete Time Evolution}
\author{\name{Atushi \surname{Tanaka}}%
  \thanks{Email address: tanaka-atushi@tmu.ac.jp}}
\inst{\address{%
    Department of Physics, Tokyo Metropolitan University,
    Hachioji, Tokyo 192-0397}}
\abst{%
  A proof of the adiabatic theorem for quantum systems whose time evolution 
  proceeds along discrete time, e.g., quantum maps and quantum circuits, 
  is shown.}

\kword{adiabatic theorem, quantum map, quantum circuit}

\newcommand{\MYVERSION}{{\bf TMU preprint} 2011-07-15}

\makeatletter
\newcommand{\CustumHeader}[1]{%
  \def\@evenhead{\footnotesize%
  \underline{%
  \hbox to \textwidth{#1\hfill}}}
  \def\ps@jpsj{\let\@oddhead\@evenhead}
}
\makeatother

\CustumHeader{\MYVERSION}

\newcommand{\set}[1]{\left\{ #1 \right\}}

\newcommand{\pdfrac}[2]{\frac{\partial #1}{\partial #2}}

\def\toexp{\mathop{\rm exp}}
\newcommand{\Texp}{\toexp_{\leftarrow}}

\newcommand{\Order}{{\mathcal O}}
\newcommand{\UK}{\hat{U}_{\mathrm{K}}}
\newcommand{\HK}{\hat{H}_{\mathrm{K}}}
\newcommand{\UD}[1]{\hat{U}_{\mathrm{D},#1}}
\newcommand{\WW}[1]{\hat{W}_{#1}}
\newcommand{\UW}[1]{\hat{U}_{{W},#1}}
\newcommand{\V}[1]{\hat{V}_{#1}}

\begin{document}
\maketitle

Quantum maps, where the time evolution proceeds along discrete time, 
describe various important models 
in studies of quantum chaos~\cite{Berry-AP-122-26}.
Also, quantum maps describe the time evolution of quantum circuits, 
which play the key role in studies of 
quantum computation~\cite{NielsenChuang}.
Many studies that concern quantum maps with slow parameters
assume that the adiabatic theorem is applicable~\cite{ApplicationOfATinQMap}.
Although the adiabatic time evolution is 
the most elementary among studies of 
quantum dynamics~\cite{Born-ZP-51-165}, 
there have been only a few works on the adiabatic theorem for
quantum maps~\cite{Berry-AP-122-26}. One is a numerical verification 
by Takami~\cite{TakamiPrivate95} and the other is
a heuristic argument by Hogg~\cite{Hogg-PRA-67-022314}. 
We note that the proof for slowly modulated 
Hamiltonians~\cite{Young-JMP-11-3298}, 
the stationary state of which is described
by an eigenvector of a Floquet Hamiltonian~\cite{Shirley-PR-138-B979}, 
is not applicable to quantum maps in general,
since the adiabatic parameter can be discontinuous in time for quantum
maps.

The aim of this note is to provide a proof of the adiabatic theorem
for quantum maps (``discrete adiabatic theorem''), where we employ 
a discrete time analog of Kato's
proof~\cite{Kato-JPSJ-5-435,MessiahAdiabaticTheorem,Avron-CMP-110-33}.
Although we will focus on the case where the spectrum of the quantum map is 
purely discrete and the spectral crossing is absent, various extensions 
should be straightforward.

We explain the adiabatic limit for quantum maps 
$\hat{U}(s)$ with a slow parameter $s$.
The adiabatic parameter evolves from $s'$ to $s''$ for $N$($>0$) steps.
Let $s_n$ be the value of $s$ at the $n$-th step ($0\le n \le N$),
where $s_0=s'$ and $s_N=s''$.
We assume that 
the interval $\delta s_n\equiv s_n - s_{n-1}$ is $\Order (N^{-1})$
as $N\to\infty$.
Also, we assume that $s_n$ belongs to a smooth path $C$. 
We will examine the time evolution induced by $\hat{U}(s)$ with 
$\set{s_n}_n$. The exact time evolution operator $\hat{U}_n$ satisfies
the recursion relation
\begin{equation}
  \hat{U}_n = \hat{U}(s_n) \hat{U}_{n-1}
\end{equation}
for $n>0$ and $\hat{U}_0=1$.

We introduce assumptions for $\hat{U}(s)$ for our proof.
First, $\hat{U}(s)$ is assumed to be unitary, and 
we assume that its spectrum consists of purely discrete eigenvalues
$\set{e^{i\theta_j(s)}}_j$, where an eigenangle $\theta_j(s)$ takes
a real value for $s\in C$.
The corresponding spectral projections $\set{\hat{P}_j(s)}_j$ satisfy
$\hat{U}(s)\hat{P}_j(s) = e^{i\theta_j(s)}\hat{P}_j(s)$ and
$\hat{P}_j(s)\hat{P}_k(s)=\delta_{jk}\hat{P}_j(s)$.
Also, we have the resolution of unity $\sum_j\hat{P}_j(s) = 1$.
The second assumption is 
\begin{equation}
  \label{eq:zCond}
  z_{jk}(s)\ne 1
  ,
  \quad\text{where}\quad
  z_{jk}(s)\equiv \exp\left(-i\{\theta_j(s)-\theta_k(s)\}\right)
  ,
\end{equation}
which corresponds to the nonzero gap condition for eigenenergies.
This also implies the absence of a crossing of eigenangles.
Hence the dimension of the $j$-th eigenspace is independent of $s$.
Finally, we assume that 
$\hat{P}_j(s)$ and $\theta_j(s)$ are smooth functions of $s$.

The discrete adiabatic theorem for quantum map $\hat{U}(s)$
is that
\begin{equation}
  \label{eq:ATinU}
  \hat{P}_j(s'')\hat{U}_N\hat{P}_k(s')
  = \delta_{jk} + \Order(N^{-1})
\end{equation}
for $N\to\infty$. 
In the following, we will follow a ``discrete analog'' of the 
conventional proof of 
the adiabatic theorem~\cite{Kato-JPSJ-5-435,MessiahAdiabaticTheorem}.
Our proof consists of two parts. One is to introduce the time evolution
operator $\WW{n}$ [eq.~(\ref{eq:defW})] for an interaction picture 
whose free evolution is the adiabatic time evolution. 
This is rather straightforward .
The other is to examine $\WW{n}$ to estimate the deviation from the
adiabatic time evolution.
In the latter part, we will frequently utilize a discrete analog of 
integration by parts:
\begin{equation}
  \label{eq:DAIP}
  \sum_{n'=1}^n (f_{n'}-f_{n'-1}) g_{n'}
  = f_n g_n - f_0 g_1 - \sum_{n'=1}^{n-1} f_{n'} (g_{n'+1}-g_{n'}) 
  .
\end{equation}

We introduce two time evolution operators that comprise 
the adiabatic time evolution.
The first part is Kato's geometric evolution operator $\UK(s, s')$,
which is assumed to satisfy the intertwining property,
\begin{equation}
  \label{eq:intertwining}
  \hat{P}_j(s)\UK(s, s') = \UK(s, s')\hat{P}_j(s')
  .
\end{equation}
$\UK(s, s')$ can be expressed by a path ordered integral of
an adiabatic ``Hamiltonian'' $\HK(s)$ along a segment of $C$:
\begin{equation}
  \label{eq:defUK}
  \UK(s, s') \equiv \Texp\left\{-i\int_{s'}^{s}\HK(r)dr\right\}
  ,
\end{equation}
where $\Texp$ represents the path ordered exponential.
Here, we employ~\cite{Kato-JPSJ-5-435}
\begin{equation}
  \label{eq:defHK}
  \HK(s)
  \equiv \frac{i}{2}\sum_j\left[\pdfrac{\hat{P}_j(s)}{s}, \hat{P}_j(s)\right]
  ,
\end{equation}
which satisfies eq.~\eqref{eq:intertwining} as well as
\begin{equation}
  \label{eq:HKdiagonal}
  \hat{P}_j(s)\HK(s)\hat{P}_j(s) = 0
  .
\end{equation}
The latter equation is convenient to prove the main theorem.
The second part of the free evolution contains 
only the dynamical phase
\begin{equation}
  \UD{n}
  \equiv \sum_j \hat{P}_j(s')
  \exp\left\{i\sum_{n'=1}^n \theta_j(s_{n'})\right\}
\end{equation}
for $n>0$ and $\UD{0} = 1$.

Using the adiabatic time evolution $\UK(s_n, s')\UD{n}$ as a
free evolution, we introduce the time evolution operator
in the interaction picture:
\begin{equation}
  \label{eq:defW}
  \WW{n}
  \equiv \left\{\UK(s_n, s')\UD{n}\right\}^{\dagger}\hat{U}_n
  .
\end{equation}
From the definition of $\UD{n}$, we have $\WW{0}=1$.
The recursion relation for $\WW{n}$ is
\begin{equation}
  \WW{n} = \UW{n}\WW{n-1}
  ,
\end{equation}
where
\begin{equation}
  \UW{n}
  \equiv
  \left\{\UK(s_n, s')\UD{n}\right\}^{\dagger}
  \hat{U}(s_n)\UK(s_{n-1}, s')\UD{n-1}
  .
\end{equation}

To prove the discrete adiabatic theorem [eq.~\eqref{eq:ATinU}], 
it suffices to show
\begin{equation}
  \label{eq:ATinW}
  \hat{P}_j(s')\WW{N}\hat{P}_k(s')= \delta_{jk}+\Order(N^{-1})
  ,
\end{equation}
which will be shown in the following.
We start from a difference equation of $\WW{n}$,
\begin{equation}
  \WW{n}-\WW{n-1} = (\UW{n}-1)\WW{n-1}
  ,
\end{equation}
which implies 
\begin{equation}
  \label{eq:Volterra}
  \WW{n} = 1 + \sum_{n'=1}^n(\UW{n'}-1)\WW{n'-1}
  .
\end{equation}
To apply the discrete analog of integration by parts [eq.~(\ref{eq:DAIP})],
we introduce 
\begin{equation}
  \V{n}
  \equiv \sum_{n'=1}^n(\UW{n'}-1)
\end{equation}
for $n>0$ and $\V{0}=0$.
From eqs.~\eqref{eq:Volterra} and \eqref{eq:DAIP}, we obtain
\begin{equation}
  \WW{n} 
  = 1 + \V{n}\WW{n-1} - \sum_{n'=1}^{n-1}\V{n'}(\UW{n'}-1)\WW{n'-1}
  .
\end{equation}
In the following, we will show 
$\UW{n}-1 = \Order(N^{-1})$ and $\V{n} = \Order(N^{-1})$, which
imply eq.~\eqref{eq:ATinW}.

We examine $\hat{P}_j(s')(\UW{n}-1)\hat{P}_k(s')$ for $j\ne k$:
\begin{align}
  \hat{P}_j(s')(\UW{n}-1)\hat{P}_k(s')
  = Z_{n-1,jk} \hat{R}_{n,jk}
  ,
\end{align}
where
\begin{align}
  \label{eq:Zdef}
  Z_{n,jk} 
  &
  \equiv \exp\left\{-i\sum_{n'=1}^n
    \left[\theta_{j}(s_{n'}) -\theta_{k}(s_{n'})\right]
    \right\}
  \intertext{and}
  \hat{R}_{n,jk}
  &
  \equiv 
  \left\{\UK(s_n,s')\right\}^{\dagger}\hat{P}_j(s_n)\hat{P}_k(s_{n-1})
  \UK(s_{n-1},s')
  .
\end{align}
From the smoothness of $\hat{P}_j(s)$,
we have
\begin{equation}
  \hat{P}_j(s_n) 
  =\hat{P}_j(s_{n-1}) + \hat{P}'_j(s_{n-1})\delta s_n + \Order((\delta s_n)^2)
  ,
\end{equation}
which implies 
\begin{equation}
  \label{eq:Restimation}
  \hat{R}_{n,jk} = \Order(N^{-1})
  \quad\text{for $j\ne k$}.  
\end{equation}
Hence, we obtain
$\hat{P}_j(s')(\UW{n}-1)\hat{P}_k(s') = \Order(N^{-1})$.
On the other hand, the ``diagonal'' part of $\UW{n}-1$ is
\begin{align}
  &
  \hat{P}_j(s')(\UW{n}-1)\hat{P}_j(s')
  \nonumber\\ &
  =  \left\{\UK(s_n,s')\right\}^{\dagger}\hat{P}_j(s_n)
  \left\{1 - \UK(s_n, s_{n-1})\right\}
  \hat{P}_j(s_{n-1})
  \UK(s_{n-1},s')
  .
\end{align}
From eqs.~\eqref{eq:defUK} and \eqref{eq:HKdiagonal}, we have
$
1 - \UK(s_n, s_{n-1})
= \Order((\delta s_n)^2)
.
$
Hence, we conclude 
$\hat{P}_j(s')(\UW{n}-1)\hat{P}_j(s')=\Order(N^{-2})$, which is
much smaller than the off-diagonal components of $(\UW{n}-1)$.

Next, we examine $\V{n}$. 
The diagonal part is 
$\hat{P}_j(s')\V{n}\hat{P}_j(s') 
= \sum_{n'=1}^n\hat{P}_j(s')(\UW{n'}-1)\hat{P}_j(s')
= \Order(n/N^2)$.
Hence, we have
$\hat{P}_j(s')\V{n}\hat{P}_j(s') =\Order(N^{-1})$
for $0 < n \le N$.
On the other hand, our estimation of the off-diagonal part
requires the destructive quantum interference 
effect induced by the dynamical phase factors to be taken into account.
Namely,
our task is to examine the following oscillatory summation,
\begin{equation}
  \label{eq:Voff}
  \hat{P}_j(s')\V{n}\hat{P}_k(s') 
  = \sum_{n'=1}^n Z_{n'-1,jk} \hat{R}_{n',jk}
  ,
\end{equation}
for $j\ne k$.
To apply eq.~(\ref{eq:DAIP}) to eq.~\eqref{eq:Voff}, we examine 
the difference of $Z_{n,jk}$:
\begin{equation}
  Z_{n,jk} - Z_{n-1,jk}
  = Z_{n-1,jk}\left\{z_{jk}(s_n) -1\right\}
\end{equation}
[see eqs.~\eqref{eq:zCond} and \eqref{eq:Zdef}].
From the noncrossing condition for the eigenangle [eq.~(\ref{eq:zCond})],
we have
\begin{equation}
  Z_{n-1,jk}
  =\frac{Z_{n,jk} - Z_{n-1,jk}}{z_{jk}(s_n) -1}
  .
\end{equation}
From eq.~\eqref{eq:Voff}, we have
\begin{align}
  \label{eq:Voff2}
  &
  \hat{P}_j(s')\V{n}\hat{P}_k(s') 
  \nonumber\\ &
  = \frac{Z_{n,jk} \hat{R}_{n,jk}}{z_{jk}(s_{n}) -1}
  - \frac{\hat{R}_{1,jk}}{z_{jk}(s_{1}) -1}
  +\sum_{n'=1}^{n-1}Z_{n',jk}\hat{R}^{(2)}_{n',jk}
  ,
\end{align}
where
\begin{equation}
  \hat{R}^{(2)}_{n,jk}
  \equiv
  \frac{\hat{R}_{n+1,jk}}{z_{jk}(s_{n+1}) -1}
  - \frac{\hat{R}_{n,jk}}{z_{jk}(s_{n}) -1}
  .
\end{equation}
The first two terms on the right-hand side of eq.~\eqref{eq:Voff2} are
$\Order(N^{-1})$ from eqs.~\eqref{eq:Restimation} and~(\ref{eq:zCond}).
It is straightforward to see that $R^{(2)}_{n}=\Order(N^{-2})$ from 
eq.~\eqref{eq:zCond} and 
the smoothness of $\hat{P}_j(s)$ and $\theta_j(s)$.
These estimations imply $\hat{P}_j(s')\V{n}\hat{P}_k(s') =\Order(N^{-1})$
for $j\ne k$.
Thus, we have confirmed eq.~\eqref{eq:ATinW}.
This completes the proof of the discrete adiabatic theorem.

\paragraph{NOTE ADDED: }
After the completion of this work, Professor Alain Joye kindly
informed me of an earlier work by A. Dranov, J. Kellendonk and
R. Seiler [J. Math. Phys. {\bf 39} (1998) 1340]. Their work is an
adaptation of ref.~\citen{Avron-CMP-110-33}, 
which is a thorough extension of Kato's proof
in ref.~\citen{Kato-JPSJ-5-435}, to discrete time evolution. 
In contrast to this, the proof
of this Short Note is a descendent of ref.~\citen{MessiahAdiabaticTheorem}, 
which is a
simplification of ref.~\citen{Kato-JPSJ-5-435} 
for systems whose spectrum is purely
discrete. It seems that a simpler argument may yet be of some value
to compare the discrete and continuous time settings.

\begin{acknowledgments}
I wish to thank Toshiya Takami for discussion.
This research was supported by the Japan
Ministry of Education, Culture, Sports, Science and Technology under
Grant number 22540396.
\end{acknowledgments}



\end{document}